\documentclass[%
 aip,
 amsmath,amssymb,
 reprint,%
]{revtex4-1}

\usepackage{graphicx}% Include figure files
\usepackage{caption}
\usepackage{dcolumn}% Align table columns on decimal point
\usepackage{bm}% bold math
\usepackage[utf8]{inputenc}
\usepackage[T1]{fontenc}
\usepackage{mathptmx}
\usepackage{etoolbox}
\usepackage{amsmath,amssymb}

\usepackage{placeins}
\usepackage{booktabs}
\usepackage{subcaption}
\usepackage{lipsum}
\usepackage{comment}
\usepackage{hyperref}
\makeatletter
\def\@email#1#2{%
 \endgroup
 \patchcmd{\titleblock@produce}
  {\frontmatter@RRAPformat}
  {\frontmatter@RRAPformat{\produce@RRAP{*#1\href{mailto:#2}{#2}}}\frontmatter@RRAPformat}
  {}{}
}%
\makeatother
\begin{document}

\preprint{AIP/123-QED}

\title[]{Benchmarking TCL4: Assessing the Usability and Reliability of Fourth-Order Approximations}
% Force line breaks with \\
\author{Jiahao Chen}
\author{Elyana Crowder}%
\author{Lian Xiang}
\author{Dragomir Davidovic}
  \email{jchen983@gatech.edu}
\affiliation{ 
Department of Physics, Georgia Institute of Technology, Atlanta, GA%\\This line break forced with \textbackslash\textbackslash
}%

\date{\today}% It is always \today, today,

\begin{abstract}
The non-Markovian dynamics of an open quantum system can be rigorously derived using the Feynman-Vernon influence functional approach. Although this formalism is exact, practical numerical implementations often require compromises. The time-convolutionless (TCL) master equation offers an exact framework, yet its application typically relies on a perturbative expansion of both the time forward and time backward state  propagators. Due to the significant computational effort involved—and the scarcity of analytical solutions for most open quantum systems—the fourth-order perturbative TCL generator (TCL4) has only been benchmarked on a limited range of systems and parameter spaces. Recent advancements, however, have made the computation of TCL4 faster and more accessible. In this paper, we benchmark the TCL4 master equation against numerically exact methods for the biased spin-boson model. We focus on the regime near critical bath coupling where perturbative master equations are expected to become inaccurate. Our findings reveal that the TCL4 approach is most reliable at low temperature and more efficient than the numerical exact methods. This study aims to delineate the conditions under which the TCL4 perturbative master equation enhances the accuracy of the TCL2.
\end{abstract}

\maketitle

\section{Introduction}
The study of open quantum systems is crucial for understanding the quantum mechanical interplay between a system and its surroundings. Exactly solvable models, such as the Jaynes-Cummings model\cite{jaynes1963comparison} and the pure dephasing regime for most models, are rare because the interaction between the bath and the system introduces significant complexity, making analytical solutions challenging, thus numerical methods become important. Numerical implementations of Feynman-Vernon influence functional formalism, directly, such as Time Evolution Matrix Product Operator (TEMPO) \cite{strathearn2018efficient,strathearn2020modelling} or indirectly, such as Hierarchical Equations of Motion (HEOM) \cite{tanimura2020numerically}, have gained popularity due to their accuracy and versatility, though each has its own compromises. In TEMPO, truncations of the memory length and data compressions with the singular value decomposition are necessary to manage the exponential growth in the dimensionality of the bath's degrees of freedom, making computations challenging when the bath correlation time is long. HEOM, on the other hand, requires the bath correlation function to be expressible as a sum of exponentials to sustain the formalism of the auxiliary density operators (ADOs). To control memory usage, the hierarchy of ADOs must be truncated. HEOM becomes cumbersome when the bath correlation function cannot be well represented by exponentials, such as in environment with algebraically decaying bath correlation functions, which occur with spectral densities that feature an exponential cutoff\cite{leggett1987dynamics}, and environment has negative noise kernel, with the low temperature limit as an example.

While those methods can be practically applied to systems with limited size across all parameter regimes, the master equation methods using the Nakajima-Zwanzig projection technique can be applied to larger systems but usually within a certain parameter regime. Over the past several decades, various master equations have been proposed under different assumptions, each applicable to specific parameter regimes \cite{davidovic2020completely, nazir2009polaron, nathan2020universal}. One of the most widely applied is the TCL master equation \cite{kubo1963stochastic, chaturvedi1979time, breuer2002theory}, which is theoretically exact and equivalent to perturbation to the Feynman-Vernon influence functional\cite{breuer2004time,nan2009nonperturbative}, but practical implementation requires a truncation of perturbative expansion. The second-order TCL(TCL2), commonly known as the Bloch-Redfield equation, has been extensively explored and used in different areas of physics, such as quantum biology\cite{ishizaki2009adequacy}, quantum transport\cite{timm2008tunneling,nestmann2019time}, quantum computing and recently the early universe in cosmology\cite{bhattacharyya2024early}. However, due to the significant numerical computation required by the original triple integral of the TCL4 generator \cite{breuer1999stochastic, breuer2001time}, benchmarking work \cite{breuer2001time, xia2024markovian,fruchtman2016perturbative,xia2024markovian,suarez2024dynamics} for non-Markovian dynamics of TCL4 has been limited. Notably, we have recently developed a fast numerical implementation of TCL4 \cite{crowder2024invalidation}, reducing the triple integral to a single integration in terms of computational complexity, thus making parameter scans of TCL4 more accessible and feasible.

This work serves two purposes: comparing the two exact methods with TCL2 and TCL4 master equations using a spin-boson model, and identifying the parameter space of temperature and bias where the accuracy of the TCL4 master equation is optimal given the limited computational resources. To determine the domain of usability of the TCL4 perturbative master equation, the coupling strength to the bath will be chosen so that the master equation is near the edge of that domain.

\section{Methods}
\subsection{Spin-boson model}
An open quantum system is described by the Hamiltonian,
\begin{equation}
    H=H_S+H_B+H_{SB}.
\end{equation}
where $H_S$, $H_B$ stand for the system and bath Hamiltonian, respectively, and $H_{SB}$ denotes the interaction between the system and bath. The spin-boson model\cite{leggett1987dynamics} describes the system as spins and the environments as bosons with different energy $\omega_k$, 
\begin{equation}
\begin{split}
    H_S=-\frac{\epsilon}{2} \sigma_z +\frac{\Delta}{2} \sigma_x&=-\frac{\Omega}{2} \sin{\theta}\sigma_z+\frac{\Omega}{2}\cos{\theta}\sigma_x,\\
    H_B&=\sum_{k}\omega_k b_k^{\dagger}b_k.
\end{split} 
\end{equation}
where $\sigma_z$, $\sigma_x$ are the Pauli matrices,  $\epsilon$ and $\Delta$ are the detuning (or bias) and tunneling parameters, respectively. The system has an energy splitting $\Omega=\sqrt{\epsilon^2+\Delta^2}$ in its eigenbasis and the system Hamiltonian can be rewritten in terms of $\Omega$ and the system bias $\theta$, where $\tan{\theta}=\epsilon/\Delta$. 

The coupling between the system and bath is mediated through the diagonal elements of spin, denoted as
\begin{equation}
    H_{SB}=\sum_{i}\frac{1}{2}\sigma_z\otimes g_i(b_i^{\dagger}+b_i).
\end{equation}
where $g_i$ are the form factors. The operators $b_i^{\dagger} (b_i)$ are creation (annihilation) operators of the boson of frequency $\omega_i$. The different choice of $\theta$ alters the relative orientation of $H_S$ and the system operator in the interaction term $\sigma_z$, ranging from parallel to orthogonal, leading to a different steady state and state dynamics. For the orthogonal case where $\theta=0$, the state is driven by the dynamics into the steady state with zero coherence, and for the parallel case $\theta=\pi/2$, the state undergoes pure dephasing with no energy transfer between the two levels. For values of 
$\theta$ between these extremes, there is a mixed effect between these two scenarios, which includes nonzero asymptotic state coherences.

The behavior of the bath is characterized by the spectral density,

\begin{equation}
    J(\omega)=\pi\sum_i |g_i|^2 \delta(\omega-\omega_i).
\end{equation}
The exact choice of spectral density is phenomenological, based on the relationship between the density of states and the frequency of bosons at low frequencies, with a cutoff term to avoid an infinite range of modes. A linear low-frequency environment corresponds to an Ohmic spectral density, which reads
\begin{equation}\label{eqn:SD}
    J(\omega)=2\pi\lambda^2 \omega f(\omega)  
    %2pi\lambda^2 \frac{2\omega_c \omega}{\omega_c^2+\omega^2}.
\end{equation}
at zero temperature, and a $f(\omega)$ is the cutoff function. Common seen cutoff are Drude-Lorentz cutoffs $f_{Drude}=\frac{\omega_c }{\omega_c^2+\omega^2}$ and exponential cutoff $f_{exp}=e^{-\omega/\omega_c}$.
At any temperature, the $J(\omega)$ of positive omega and negative omega are associated through the Kubo-Martin-Schwinger (KMS) condition\cite{kubo1957statistical}, which enforces a discontinuity on the derivative of the cutoff function of the spectral density at zero frequency if the derivative is not zero at zero frequency such as exponential cutoff. To get the time-dependent spectral density, one first needs the bath correlation function (BCF) 
\begin{equation}\label{eqn:bcf}
C(t)=\frac{1}{\pi}\int_{0}^{\infty}d\omega'  J(\omega')\left[\cos{\omega't}\coth{\frac{\hbar\omega'}{2k_BT}}-i\sin{\omega't}\right],
\end{equation} 
where the real part is the noise kernel with temperature dependence and the imaginary part the dissipation kernel. Here $T$ is the temperature and $\hbar, k_B=1$ is assumed in this paper. For an spectral density in Eqn. \ref{eqn:SD} with $f_{Drude}$, the real correlation function become negative and non-Makovian at early time at low temperature; for an exponential cutoff $f_{exp}$, the correlation function decays algebraically rather than exponentially at long time scale. 

With the BCF, the timed spectral density is computed as
\begin{equation}\label{eqn:time_sd}
    \Gamma(\omega,t)=\int_{0}^{t} C(t')e^{i\omega t'} dt'.
\end{equation}
\subsection{Time convolutionless master equation}

The free dynamics of system and bath are easy to solve, and we can move to the interaction picture for both the ease of derivation of the master equation and the need to treat the interaction as perturbation when necessary. The density matrix evolution in the interaction picture is described by Louiville-von Neumann equation, 
\begin{equation}
    \frac{d\rho_t(t)}{dt}=\mathcal{L}(t) \rho_t(t)\label{eqn:von-Neumann}
\end{equation}
where $\rho_t(t)$ stands for the density matrix of the product Hilbert space of system and bath $\mathbb{H_S}\otimes \mathbb{H_B}$, and the superoperator $\mathcal{L}(t)[\boldsymbol{\cdot}]=-i[H_{SB}, \boldsymbol{\cdot}]$. 

TCL generators are derived explicitly from the Eqn \ref{eqn:von-Neumann} by applying the Nakajima-Zwanzig technique, which defines the relevant part
\begin{equation}
    \mathcal{P}(\rho)=tr_B(\rho)\otimes \rho_B,
\end{equation}
and the irrelevant part,
\begin{equation}
    \mathcal{Q}=1-\mathcal{P}.
\end{equation}

By taking the relevant and irrelevant parts of Eqn.(\ref{eqn:von-Neumann}), one can get the equations
\begin{equation}
    \frac{d}{dt}\mathcal{P}\rho(t)= \mathcal{P}\mathcal{L}(t)\mathcal{P}\rho(t)+\mathcal{P}\mathcal{L}(t)\mathcal{Q}\rho(t)\label{eqn:relevant part}
\end{equation}
\begin{equation}
    \frac{d}{dt}\mathcal{Q}\rho(t)= \mathcal{Q}\mathcal{L}(t)\mathcal{P}\rho(t)+\mathcal{Q}\mathcal{L}(t)\mathcal{Q}\rho(t)\label{eqn:irrelevant part equation}
\end{equation}

Solving Eqn. (\ref{eqn:irrelevant part equation}), and plugging the formal solution into Eqn. (\ref{eqn:relevant part}), we reach the Nakajima-Zwanzig generalized master equation for the system,
\begin{equation}
    \frac{d}{dt}\rho_S(t)=\mathcal{P}\mathcal{L}(t)\rho_S(t)+\int_0^t ds  \mathcal{P} \mathcal{L}(t) \mathcal{G}(t,s)\mathcal{L}(s)\rho_s(s)
\end{equation}
where the inhomogeneous part is neglected given the factorized initial condition $\rho=\rho_S\otimes \rho_B$, throughout in the paper. $\rho_s(t)$ is the reduced density matrix of system at time $t$. Here, $\mathcal{G}(t,s)=T_{\leftarrow} e^{\lambda \int_s^t \mathcal{Q} \mathcal{L}(s') ds' }$ and $T_{\leftarrow}$ is the time-ordering operator.

One has to calculate the differential-integral equation at each timepoint to propagate the system density matrix. To make the equation time-local,
the TCL master equations use a back-order propagator to transfer the time nonlocality in the memory kernel of the system to the bath function. 
\begin{equation}
        \frac{d}{dt}\rho_S(t)=\mathcal{L}(t)\rho_S(t).
\end{equation}
where

\begin{equation}
    \begin{split}
    \mathcal{L}(t)=\lambda \mathcal{P}\mathcal{L}(t)[1-\Sigma(t)]^{-1}\mathcal{P}\\
    \Sigma(t)=\lambda\int_0^t ds \mathcal{G}(t,s)\mathcal{Q}\mathcal{L}(s)\mathcal{P}G(t,s).
    \end{split}
\end{equation}
where $G(t,s)=T_{\rightarrow} e^{-\lambda \int_s^t \mathcal{Q} \mathcal{L}(s') ds' }$.
The TCL generators can be derived by expanding $[1-\Sigma(t)]^{-1}$ into a geometric series, with each odd order $\mathcal{L}_{2n+1}(t)$ eliminated under a Gaussian bath assumption, and only even terms $\mathcal{L}_{2n}(t)$ left. The fourth-order master equation is written as 
\begin{equation}\label{eqn:master1}
    \frac{d\rho(t)}{dt}=\mathcal{L}(t)\rho(t)=[\mathcal{L}_0(t)+\lambda^2\mathcal{L}_2(t)+\lambda^4\mathcal{L}_4(t)]\rho(t),
\end{equation}

The combination of zeroth order and second order provides the widely known Redfield generator (or TCL2 generator interchangeably). The generators $
\mathcal{L}_0$ and $\mathcal{L}_{2}$ in matrix form is given by  

\begin{equation}
    \begin{aligned}
\mathcal{L}_{nm,ij}^{(0)}&=-i[[E_{n}-E_{m}]\delta_{ni}\delta_{mj},\\
\mathcal{L}_{nm,ij}^{(2)}&=\{A_{ni}A_{jm}[\Gamma_{in}(t)+\Gamma_{jm}^{\star}(t)]\\
&-
\sum_{k=1}^{N}[A_{nk}A_{ki}\delta_{jm}\Gamma_{ik}+\delta_{ni}A_{jk}A_{km}\Gamma_{jk}^{\star}]\},
\end{aligned}
\end{equation}

The expansion to the fourth order yields a time-ordered triple integration over time shown as Eqn (29) in the paper \cite{breuer2001time}. The detailed derivation leads to the simplified bath operator can be seen from Crowder et al \cite{crowder2024invalidation}. We assume that the system coupling operator is time-independent in the Schrodinger picture so that the system operator is separated from the integration, leaving one integral of the triple integral to be effectively expressed by the timed spectral density $\Gamma_\omega(t)$, which can be pre-calculated. The further integral simplification\cite{crowder2024invalidation} expresses the generator into the bath functions $\mathsf{F}$, $\mathsf{C}$, $\mathsf{R}$, of which the compuation requires a convolution in $\mathsf{F}$ and $\mathsf{C}$ and a simple integration in $\mathsf{R}$. The fourth order generator is

\begin{equation}
\begin{aligned}
       \mathcal{L}_{nm,ij}^{(4)}(t) &=\sum_{a,b,c=1}^N\Big\{     
        A_{na}A_{ab}A_{bc} A_{ci} \delta_{jm}[\mathsf{F}_{cb,ci,ac}(t) \\
        &- \mathsf{R}_{cb,ab,bi}(t)]
        + A_{na} A_{ab} A_{bc}A_{ci} \delta_{jm}\mathsf{R}_{ic,ab,bi}(t)
        \\
        &- A_{na}A_{ab} A_{bc}A_{ci}\delta_{jm}\mathsf{F}_{ba,ci,ac}(t) \Big\}\\
        &\sum_{a,b}^N\Big\{-A_{na} A_{ab} A_{bi} A_{jm}  [\mathsf{F}_{ba,bi,nb}(t) - \mathsf{R}_{ba,na,ai}(t)]
        \\
        &+ A_{na} A_{ab} A_{bi} A_{jm} \mathsf{F}_{an,bi,nb}(t)\\
        &- A_{na} A_{ab}A_{bi}A_{jm}\mathsf{R}_{ib,na,ai}(t)
        \\
        &+ A_{na} A_{ab} A_{bi}A_{jm} [\mathsf{C}_{ba,jm,ai}(t)+\mathsf{R}_{ba,jm,ai}(t)]\\
        &- A_{na} A_{ab} A_{bi} A_{jm}[\mathsf{C}_{ib,jm,ai}(t)+\mathsf{R}_{ib,jm,ai}(t)]
        \\
        &- A_{na} A_{ai} A_{jb}A_{bm}[\mathsf{C}_{an,jb,ni}(t)+\mathsf{R}_{an,jb,ni}(t)]\\
        &+ A_{na} A_{ai} A_{jb}A_{bm} [\mathsf{C}_{ia,jb,ni}(t)+\mathsf{R}_{ia,jb,ni}(t)]\Big\}+R.H.C,
\end{aligned}
\end{equation}
where $R.H.C(\mathcal{L}_{nm,ij}^{(4)}(t))=L.H.C({\mathcal{L}_{ij,nm}^{(4)}}^{\star}(t))$

The bath operators $\mathsf{F}$, $\mathsf{C}$ and $\mathsf{R}$ are defined as
\begin{equation}
    \begin{aligned}
    \mathsf{F}_{\omega_1\omega_2\omega_3}(t) &=-\int_0^{t} d\tau \Delta \Gamma_{\omega_1}(t,\tau) \Delta \Gamma_{\omega_2}^{ T}(t,t-\tau) e^{-i(\omega_1 + \omega_2 + \omega_3)\tau}\\
    &+i\Gamma_{\omega_2}^{\beta T}(t)\frac{\Gamma_{-\omega_2-\omega_3}(t)-\Gamma_{\omega_1}(t)}{\omega_1+\omega_2+\omega_3}\\
    \mathsf{C}_{\omega_1\omega_2\omega_3}(t) 
    &=-\int_0^{t} d\tau \Delta \Gamma_{\omega_1}(t,\tau) \Delta\Gamma_{\omega_2}^{\star}(t,t-\tau) e^{-i(\omega_1
     + \omega_2 + \omega_3)\tau}\\
&+i\Gamma_{\omega_2}^{\star}(t)\frac{\Gamma_{-\omega_2-\omega_3}^\alpha(t)-\Gamma_{\omega_1}}{\omega_1+\omega_2+\omega_3}\\
   \mathsf{R}_{\omega_1\omega_2\omega_3}(t) 
   &=-\int_0^{t} d\tau \Delta \Gamma_{\omega_1}(t,\tau) \Delta \Gamma_{\omega_2}(t,\tau) e^{-i(\omega_1 + \omega_2 + \omega_3)\tau}\\
   &+i\Gamma_{\omega_2}(t)\frac{\Gamma_{-\omega_2-\omega_3}(t)-\Gamma_{\omega_1}(t)}{\omega_1+\omega_2+\omega_3},
    \end{aligned}
\end{equation}
and
\begin{equation}
    \begin{aligned}
    \Delta \Gamma_{\omega}(t,s) = \Gamma_{\omega}(t) - \Gamma_{\omega}(s).\\
\end{aligned}
\end{equation}
The overall computational complexity to get the TCL4 generator at time $t$ is as complex as a single integration. However, the integrals in F and C  are convolutions that must be computed from 0 to $t$ at each $t$.
%%
%We assume that the system coupling operator is time-independent in the Schrodinger picture so that the system operator is separated from the integration, leaving one integral of the triple integral to be effectively expressed by the timed spectral density $\Gamma_\omega(t)$, which can be pre-calculated. The further integral simplification\cite{crowder2024invalidation} expresses the generator into the bath function $\mathsf{F}$, $\mathsf{C}$, $\mathsf{R}$, which requires a single integration. The overall computational complexity to get the TCL4 generator is as complex as a single integration. 
%%

\begin{figure*}
\captionsetup{justification=raggedright,singlelinecheck=false}
\includegraphics[width=\linewidth]{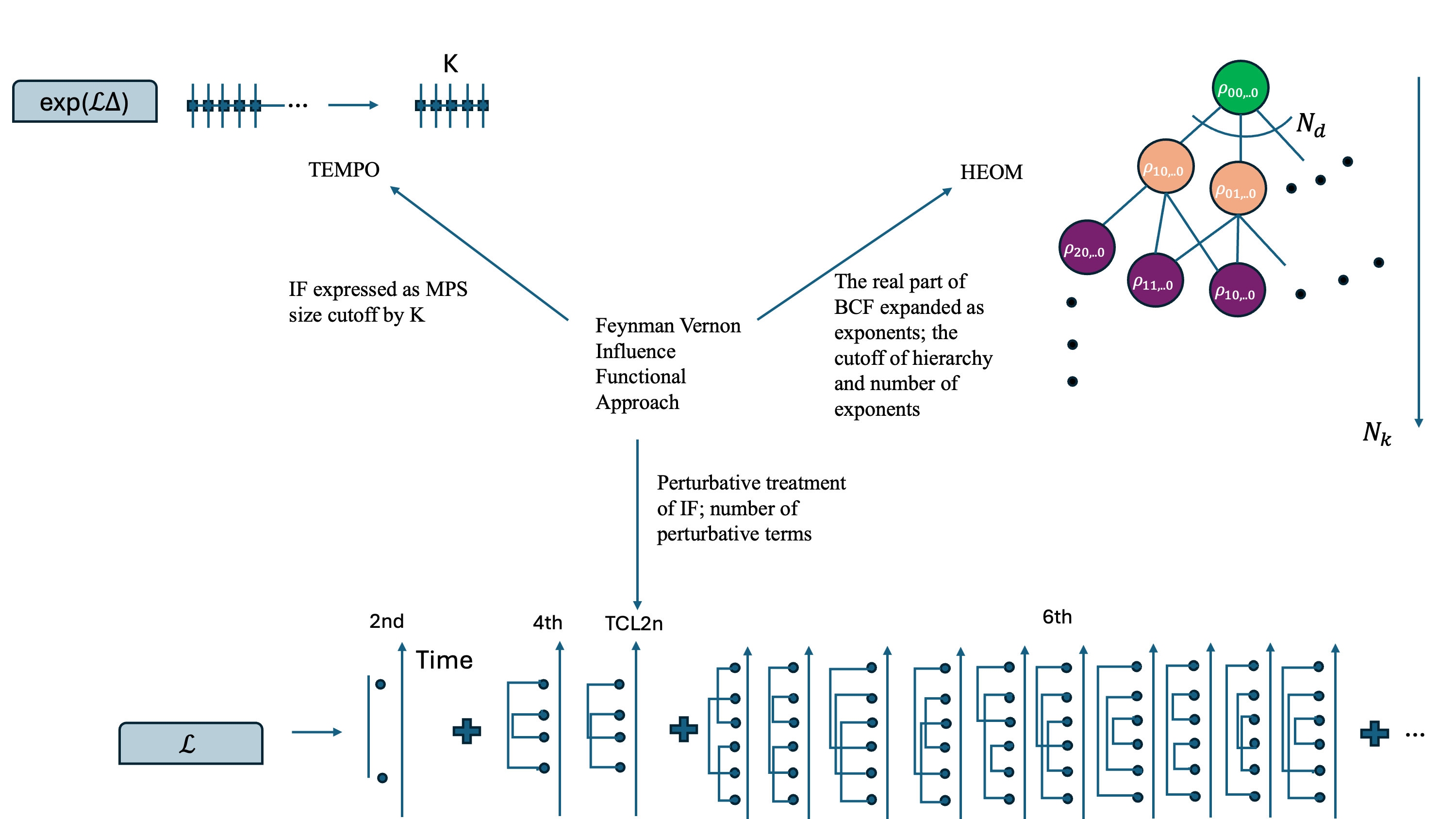}   \caption{Comparison of TCL, TEMPO, and HEOM, all of which may be obtained using the influence functional (IF) technique.
TEMPO  represents  the IF as a Matrix Product State (MPS), with the bath memory time limited by $K\Delta_t$. HEOM generates auxiliary density operators (ADOs) based on IF's density matrix derivatives. The number of bonds in the first density matrix corresponds to the number of bath modes $N_d$, and a hierarchy truncation of $N_k$ is applied. TCL expresses the IF as a perturbative expansion in the coupling strength. The Liouvillian is  obtained by combining  the backward and the derivative of the forward propagator, and is also perturbatively expanded and local in time.}
    \label{fig:enter-label}
\end{figure*}
\subsection{HEOM}
The HEOM method is derived from the Feynman-Vernon influence functional formalism through intergration by parts, which only works if the noise and dissipation kernels are exponential functions\cite{tanimura2020numerically}. This method reformulates the system-environment interaction by expanding it into a hierarchy of auxiliary density operators (ADOs), where the bath correlation functions are expressed as a sum of exponentials decay modes. Within this assumption, the ADOs are associated with the nearest ADOs by the differential equations. To ensure convergence and accuracy in HEOM simulations, several key parameters must be tuned. The hierarchy depth $N_k$ controls the number of ADOs included, with larger $N_k$ capturing higher-order system-bath correlations but increasing computational costs. The number of exponential terms $N_{d}$ 
determines how accurately the bath correlation function is approximated, where larger $N_{d}$ improves precision but demands more resources. Therefore, a bath correlation function that cannot be accurately expressed in exponential modes poses a challenge for HEOM. Examples include low-temperature limits, where the non-Markovian negative noise kernel is difficult to express. Another example is a spectral density with an exponential cutoff, where the bath correlation function exhibits algebraic decay at large times. In this work, we use HEOM solver in the Qutip package\cite{johansson2012qutip,lambert2023qutip} to perform HEOM computation. 
\subsection{TEMPO}
Our TEMPO simulations utilize open-source software detailed in paper by Strathearn et.al \cite{strathearn2018efficient}. TEMPO employs matrix product product state/operator representation \cite{pirvu2010matrix, orus2014practical} to simulate the dynamics of open quantum systems, relying on singular value decomposition and the truncation of small singular values to reduce tensor dimensionality. The precision parameter 
$\lambda_c$
  sets the threshold for singular values retained in each decomposition, while the time evolution is dependent on Trotterization and discretization with a time step 
$\Delta_t$. To capture the non-Markovian influence of the bath on the system, TEMPO propagates an augmented density tensor—a tensor documenting the possible system propagation choices from one time step to all subsequent steps—and retains $K$ tensor legs at each time step to prevent exponential memory growth. 
Despite the TEMPO method being theoretically exact, the memory cutoff ($K$), precision parameter ($\lambda_c$), and time step ($\Delta_t$) introduce inaccuracies into the computation. Although it is not possible to evaluate the effects of these parameters a priori, we empirically determine the confidence level of different parameters used in the simulation. Generally, we decide if the TEMPO results are trustworthy by observing the convergence when varying these parameters. However, in some cases, higher accuracy requirements (increasing $K$ or $\lambda_c$) can lead to unstable data points in TEMPO. The goal of this article is to identify the validity region of TCL4. To achieve this, TEMPO parameters are considered acceptable if difference of TEMPO results of two sets of parameters exhibit variations much smaller than the discrepancies observed between TCL4 and TCL2. More specifically, if the parameter controlling convergence is increased, and the trace distance of the density matrix (computed between the updated TEMPO parameters and the previous ones) is significantly smaller than the trace distance observed with TCL4 and TCL2, the new TEMPO parameters are finally adopted.

\section{Results}
\begin{figure*}
\captionsetup{justification=raggedright,singlelinecheck=false}
    \centering
\includegraphics[width=\linewidth]{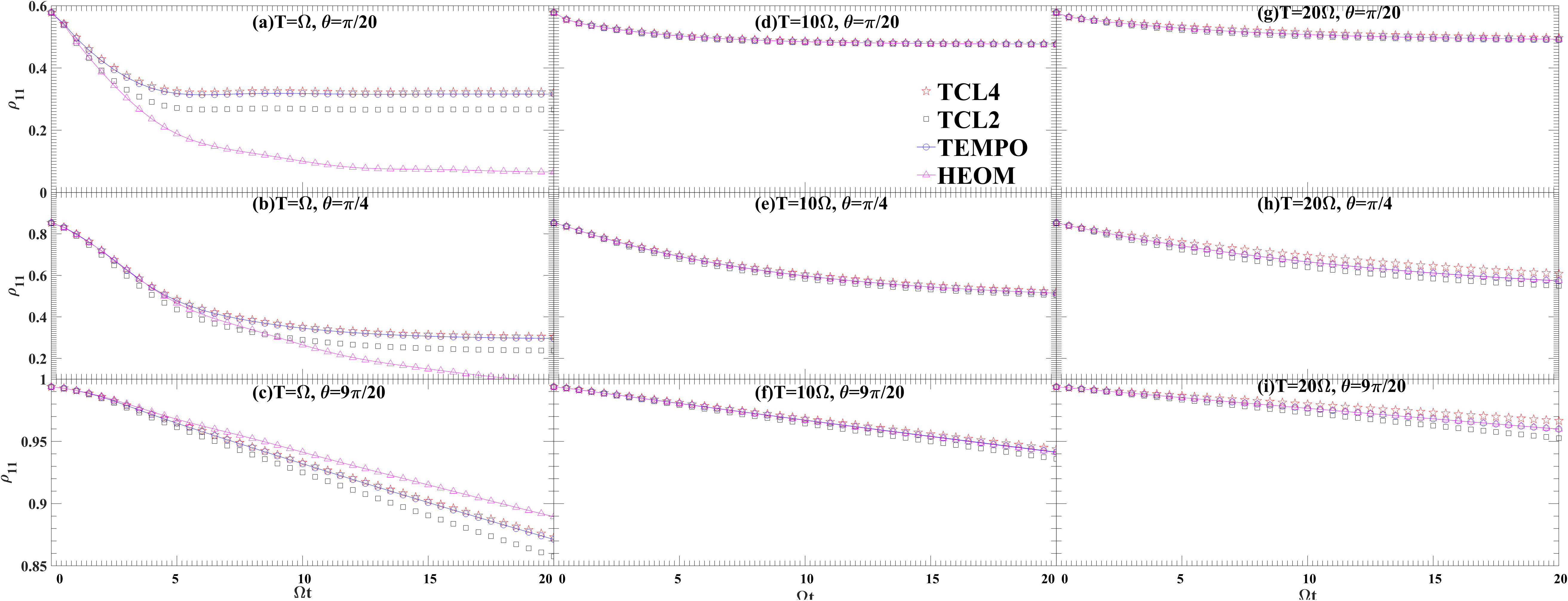}
    \caption{Population dynamics at different temperatures and bias. (a-c) $T=\Omega$, $\theta=\pi/20$, $\pi/4$, $9\pi/20$. (d-e) $T=10\Omega$, $\theta=\pi/20$, $\pi/4$, $9\pi/20$.
    (g-i) $T=20\Omega$, $\theta=\pi/20$, $\pi/4$, $9\pi/20$. Model parameters: Simulation parameters: (TEMPO) $\Delta_t=0.01$, $\lambda_c=75$, $K=1000$; (HEOM) $\Delta_t=0.01$, $N_k=8$, $N_d=8$. $f_w=\frac{\omega_c}{\omega_c^2+\omega^2}$ (Drude-Lorentz Cutoff). 
}
    \label{fig:Fig1}
\end{figure*}
\begin{figure*}
    \centering
\captionsetup{justification=raggedright,singlelinecheck=false}
\includegraphics[width=\linewidth]{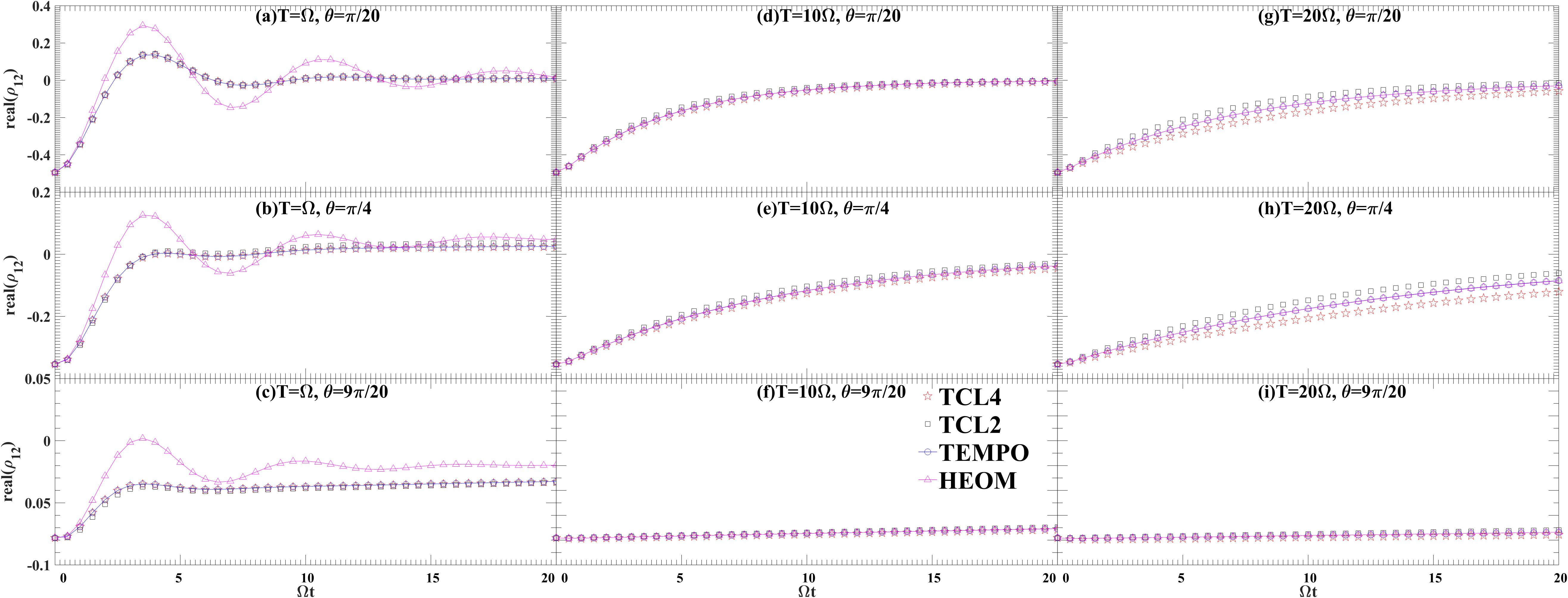}
    \caption{Coherence dynamics at different temperatures and bias, corresponding to Fig. \ref{fig:Fig1} (a-i). 
    }\label{fig:Fig2}
\end{figure*}
\subsection{\textit{Population and Coherence Dynamics}}
Our benchmarking explores variations in both the bath temperature and the system bias, denoted by $\theta$. We begin by discussing the population and coherence dynamics for three different bias scenarios: small bias ($\theta = \pi/20$), medium bias ($\theta = \pi/4$), and large bias ($\theta = 9\pi/20$). Throughout this comparison, we assume an Ohmic bath with a cutoff frequency of $\omega_c = 10\Omega$ and a coupling strength $\lambda^2 = 1$. The norm ratio of the fourth-order term to the second-order term is
usually less than 0.1, so the coupling strength is relatively strong. The temperatures considered are $T = \Omega$, $10\Omega$, and $20\Omega$. Starting from this section, the initial states are chosen as
\begin{equation}
    \rho_0=V^\dagger \begin{pmatrix}
        1&0\\
        0&0
    \end{pmatrix}V
\end{equation}
in the eigenbasis of $H_S$, and 

\begin{equation}
    V=\frac{1}{\sqrt{2}}\begin{pmatrix}
    -\sqrt{\cos{\theta}+1}&\sqrt{1-\sin{\theta}}\\
    \sqrt{1-\sin{\theta}}&\sqrt{\cos{\theta}+1}
\end{pmatrix}
\end{equation} has the eigenvectors of $H_S$. V transforms the $H_s$ into the eigenenergies $E=V'H_S V=\begin{pmatrix}
    \Omega/2 & 0\\
    0& -\Omega/2
\end{pmatrix}$. The initial states varing with the bias then is
\begin{equation}
    \rho_0(\theta)= \frac{1}{2} \begin{pmatrix}
        \sin{\theta}+1&-\cos{\theta}\\
        -\cos{\theta}& -\sin{\theta}+1
    \end{pmatrix}
\end{equation}

Figure \ref{fig:Fig1} and Figure \ref{fig:Fig2} demonstrate the population and coherence dynamics at different temperatures and biases. 
As bias and temperature ($T$) increase, the relaxation time of the system also increases. Specifically, at $T=\Omega$, $\theta=\pi/20$ and $\pi/4$, the population exhibits weak oscillations, whereas the population of $\theta=9\pi/20$ does not show an oscillation, indicating a transition from coherent transport to incoherent transport along $\theta$. As
$\theta$ increases, the system's Hamiltonian $H_S$ becomes more aligned with the system coupling operator $\sigma_z$, resulting in an increase in the dephasing rate and a reduction in the overall coherence of the system. At intermediate and high temperatures, population transport from the excited state becomes predominantly incoherent irrespective of bias as higher temperatures suppress the coherence by increasing the dephasing rate. Meanwhile, the temperature drives the steady-state upward across all three biases due to the higher bath correlation function. However, the relaxation rate decreases. In weak coupling theory, an increasing relaxation rate is typically observed as the temperature increases. This phenomenon is discussed in the Appendix \ref{appendix: relexation}.

At low temperatures, HEOM faces significant challenges and shows poor agreement with TEMPO, despite aligning well with TEMPO at higher temperatures. HEOM relies on a finite set of basis functions, typically decaying exponentials, but at low temperatures, it requires substantially more exponential bath terms, making it computationally impractical. This issue stems from the scale-free noise spectrum at low frequencies, which is associated with zero-point quantum fluctuations. Although advances have aimed to address the limitations of HEOM in the low temperature regime by either including the correction terms or using curve fitting to shrink the bath terms needed\cite{fay2022simple,moix2013hybrid,tang2015extended,tanimura2020numerically}, here we use the traditional approach.

In contrast, the TCL2n master equation maintains a constant computational cost regardless of temperature. TCL4 is a result-reliable and computationally effective option at low temperatures, showing excellent consistency with TEMPO, especially for both population and coherence dynamics, as illustrated in Figures \ref{fig:Fig1} (a)-(c) and \ref{fig:Fig2} (a)-(c). For example, the population elements between TCL4 and TEMPO differ by approximately $2\%$ at $T = \Omega$, $\theta = \pi/20$, whereas for TCL2, this difference is $15\%$. Regarding the coherences at $T=\Omega$, TCL4 always aligns closely with TEMPO, and TCL2 also aligns with TEMPO at low bias, but starts to deviate from TEMPO at $\theta=\pi/4$ and $9\pi/20$.  TCL4 effectively remedies the shortcomings of TCL2, increasing accuracy and correcting the equilibrium state at low temperatures.

Although TEMPO remains exact at low temperatures, its computational complexity grows exponentially with the simulation time, unlike the linear scaling seen with the HEOM and TCL methods. Furthermore, long-time simulations with TEMPO at low temperatures can introduce numerical instabilities, complicating its use for extended simulations.

As the temperature increases to $T = 10\Omega$, the HEOM shows better agreement with TEMPO, as expected for this regime. For all the three biases, TEMPO and HEOM have around 0.01\% difference of populations. Compared to the case of the low temperature limit, at $T=10\Omega$, the behavior of TCL4 against TCL2 diverges at different biases. In the case of TCL2, the population dynamics begins to match more closely the exact method at $10\Omega t$ compared to the low temperature limit. TCL4 remains more accurate than TCL2 at low and medium bias levels, but begins to lose reliability at higher biases, where the population shows notable deviations from the correct results. This trend becomes even more evident at higher temperatures. At $T = 20\Omega$, TCL4 is only better than TCL2 in population dynamics at $\theta = \pi/20$. For other populations and coherence dynamics at different biases, TCL4 is as erroneous as TCL2, though they deviate on opposite sides of the TEMPO curve due to the perturbative nature of the TCL equations. This suggests that TCL4 becomes unusable at excessively high temperatures.

\subsection{Trace Distance of Population and Coherence}
\begin{figure}
\captionsetup{justification=raggedright,singlelinecheck=false}
\includegraphics[width=\linewidth]{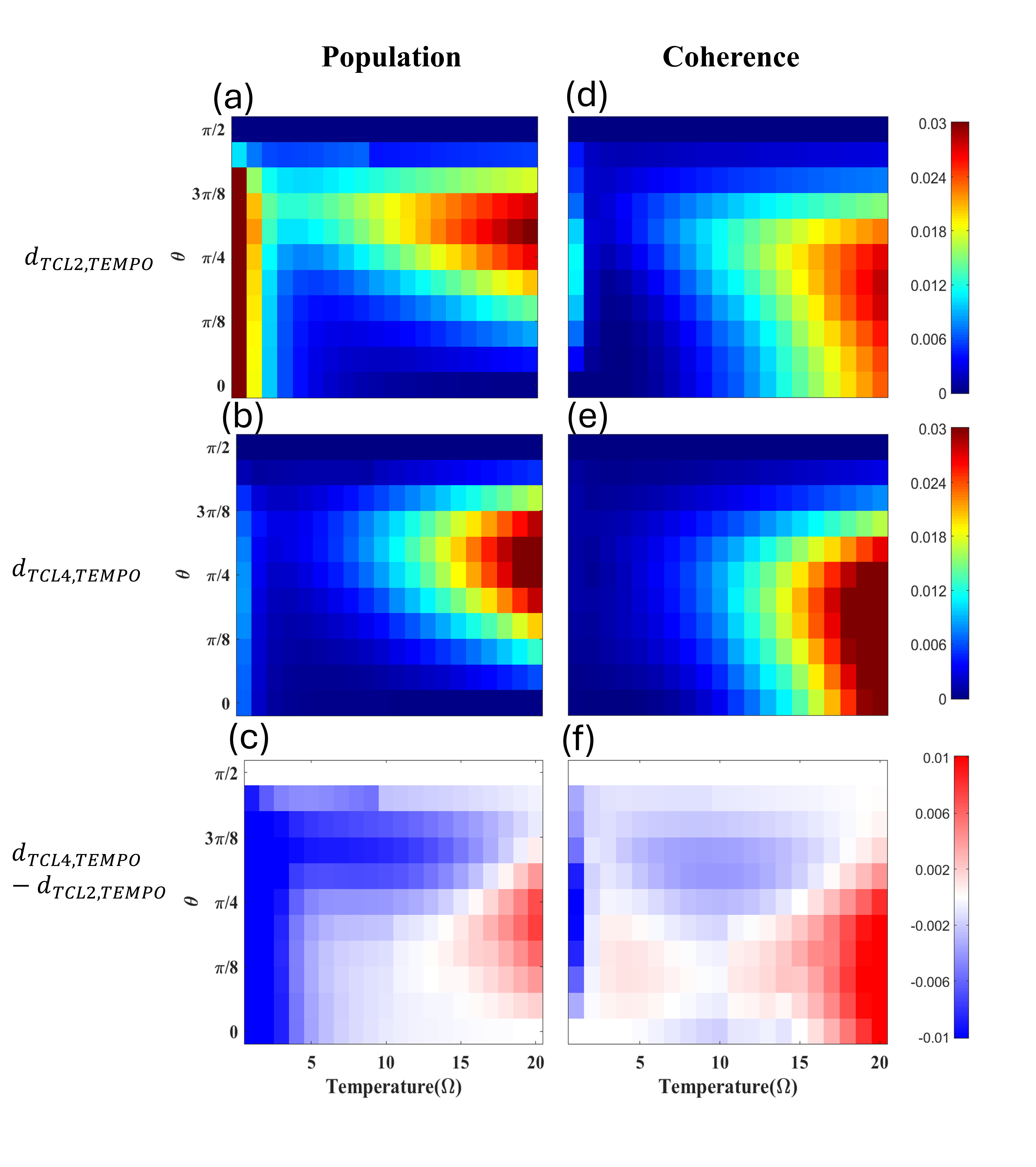}
    \caption{The trace distance between TCLs and TEMPO, population and coherence separately. (a-c) Population. (d-e) Coherence. $t=15/\Omega$. }
    \label{fig:coh_pop}
\end{figure}
To quantify the impact of the fourth order term on the solution to the TCL equation in different parameter regimes, we measure the trace distance at some time point t,
\begin{equation}
    d_{\rho_1,\rho_2}(t)=\frac{1}{2}\|\rho_{1}(t)-\rho_{2}(t)\|_1
\end{equation}
with $\rho_1$ being the density matrix of one method and $\rho_2$ for the other method. $d_{\rho_1,\rho_2}(t)=0$ means that the density matrix computed by the two methods is identical at time t, and $d_{\rho_1,\rho_2}(t)=1$ means that they are completely orthogonal and distinguishable.

Figure \ref{fig:coh_pop} presents the trace distance for coherence and population elements between TCL2, TCL4, and TEMPO at
$t=10$. As shown in Figures \ref{fig:coh_pop} a and \ref{fig:coh_pop} b, TCL2 exhibits significant errors in population dynamics at low temperatures, 
$\theta=0$ to $\theta=9\pi/20 $, while TCL4 maintains much smaller errors. In contrast, TCL2 shows smaller errors in coherence dynamics, while the error in TCL4's coherence dynamics remains below $1\times 10^{-4}$. In general, both TCL2 and TCL4 perform better in coherence dynamics than in population dynamics at low temperatures. As the temperature increases, the error in population dynamics for both TCL2 and TCL4 starts to increase at intermediate bias values, although they continue to converge with TEMPO at zero and extreme bias. In contrast, the coherence dynamics follow a different trend, with the error increasing from zero bias to
$\theta=3\pi/8$, as in Figure \ref{fig:coh_pop} (d) and (e). In Figures \ref{fig:coh_pop} (c) and \ref{fig:coh_pop} (f), TCL4 does not consistently outperform TCL2. As the temperature exceeds $15\Omega$, TCL4 begins to perform worse than TCL2 in both coherence and population dynamics.
\subsection{Time-Averaged trace distance}
While the trace distance gives a snapshot of how distinguishable two quantum states are at a particular moment, the time-averaged trace distance measures their average distinguishability over a specified period. This captures the overall behavior of the two states across the time interval
\begin{equation}
    \Delta_{\rho_1,\rho_2}=\frac{1}{t_{end}}\int_{0}^{t_{end}}d_{\rho_1,\rho_2}(t) dt,
\end{equation}
where $t_{end}$ is the cutoff time for calculating the averaged trace distance.

\begin{figure}
\captionsetup{justification=raggedright,singlelinecheck=false}
\includegraphics[width=\linewidth]{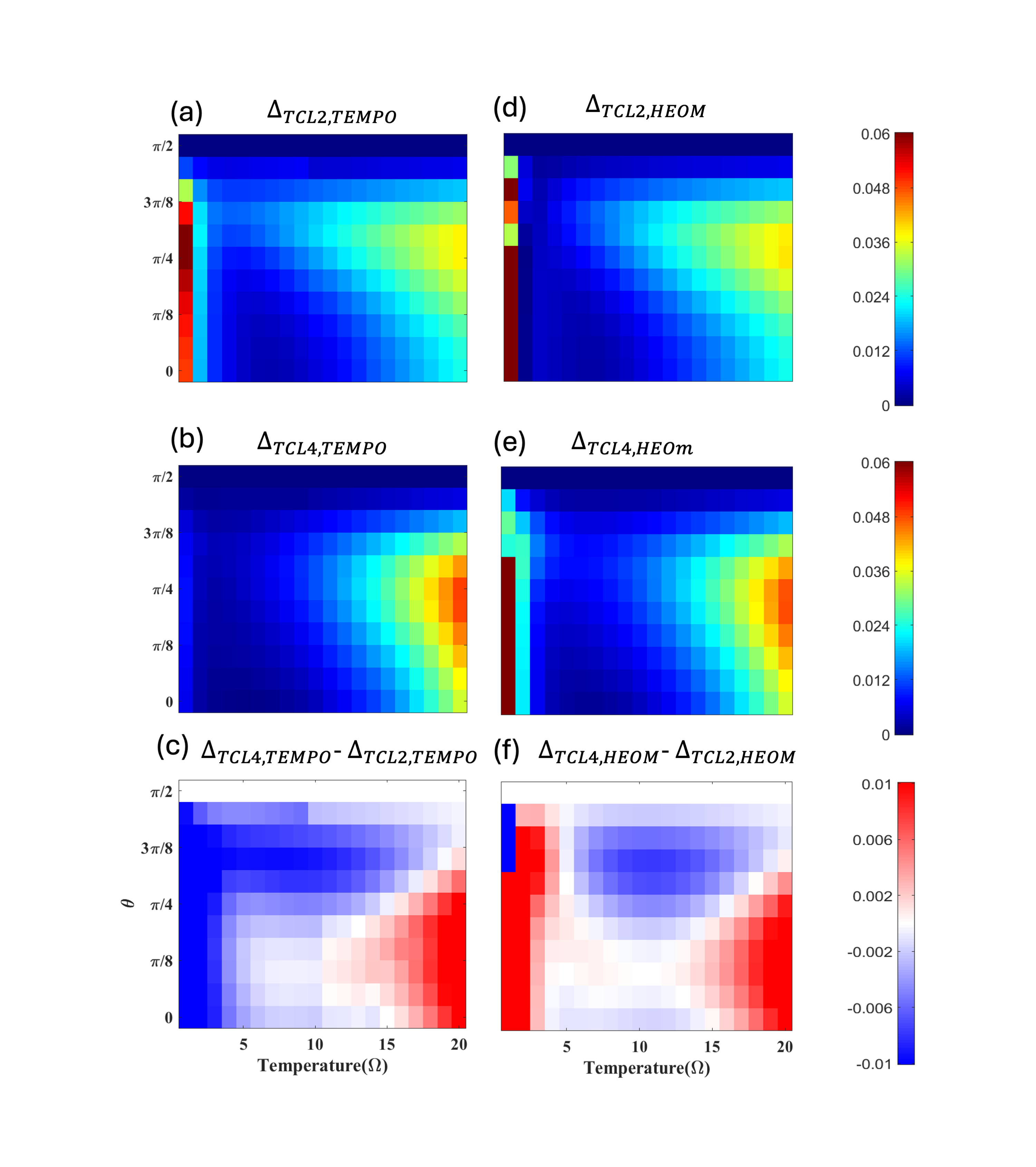}
    \caption{The time-averaged trace distance between TCLs and exact methods. Convergence parameters: (TEMPO) $T=\Omega$, $\Delta_t=0.01$, $\lambda_c=80$, $K=1000$. For other temperatures, $\Delta_t=0.01$, $\lambda_c=75$, $K=1000$. (HEOM) $N_k=8$. $N_d=8$. }
    \label{fig:timeaverate}
\end{figure}

Figures \ref{fig:timeaverate} (a) and (b) show the average time-averaged trace distance between TCLs and TEMPO. As in the previous section, TCL4 is significantly more accurate at low temperature. The time-averaged trace distance between TCL2 and TEMPO is greater than 0.02 from $\pi=0$ to $3\pi/8$, while the error of TCL4 is below 0.005. The error of TCL2 approaches TCL4 when increasing $\theta$, and at $\theta=\pi/2$, the errors of TCL2 and TCL4 become equivalent, as shown in Figure \ref{fig:timeaverate} (c). This is a known result that in the pure dephasing regime, where $H_s$ is parallel to $\sigma_z$, the second-order TCL equation can provide an exact solution. From low temperatures to $T=5\Omega$, the averaged trace distance of TCL2 generally decreases to values below 0.02. In this temperature region, TCL4 consistently outperforms TCL2. However, beyond $T=5\Omega$, the error in TCL2 increases as the temperature continues to rise while TCL4 shows a similar trend. Notably, in \ref{fig:timeaverate} (c), as the temperature is beyond $T=19\Omega$, a clear boundary emerges where TCL4 begins to perform worse than TCL2. In contrast to HEOM, high temperatures are the critical limitation for TCL4, whereas, at low temperatures, TCL4 remains a reliable and usable method for systems coupled to an Ohmic bath. 

\begin{figure}
\captionsetup{justification=raggedright,singlelinecheck=false}
  \begin{subfigure}{0.49\columnwidth}
  \includegraphics[width=\textwidth]{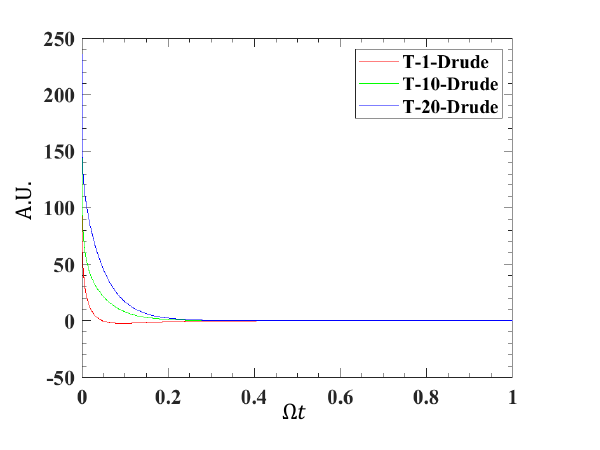}
  \caption{}
  \end{subfigure}
  \hfill
  \begin{subfigure}{0.49\columnwidth}
  \includegraphics[width=\textwidth]{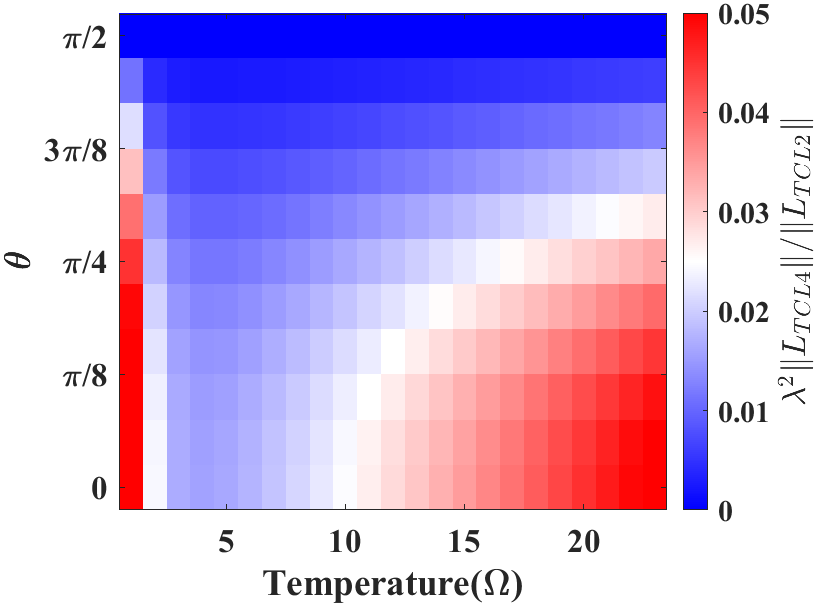}
  \caption{} 
  \end{subfigure} 
  \caption{(a)The fluctuation (real part of BCF) at different temperature. (b) The norm ratio of fourth order generator and second order generator. $\lambda^2\frac{\|\mathcal{L}_4\|}{\|\mathcal{L}_2\|}$ at $t=15/\Omega $. }\label{fig:norm_bcf}\end{figure}

The two parameter regions where TCL2 exhibits significantly higher error compared to other regions are the low-temperature limit and the high-temperature regime with increased bias. This discrepancy is reflected in the norm ratio, as shown in Figure \ref{fig:norm_bcf} (b), which presents the norm ratio between the fourth-order and second-order generators as an indicator of the failure of perturbative expansion. The regions where the norm ratio peaks correspond closely to those where the time-averaged trace distance between TCL2 and TEMPO becomes prominent, as seen by comparing Figure \ref{fig:timeaverate} (a) with Figure \ref{fig:norm_bcf} (b). The breakdown of TCL2 at low temperatures is attributed to positivity violations due to the negative noise kernel at early times; TCL4 substantially mitigates these positivity violations. Furthermore, positivity violation primarily impacts the population in the density matrix, as illustrated in Figure \ref{fig:coh_pop} (a). At high temperatures, invalidity arises from coherence elements, with the fourth-order generator improving accuracy up to
$T<20\Omega$; however, TCL4 itself becomes unreliable beyond this point.

The only term that introduces temperature into the computation is the hyperbolic function $\coth{\frac{\omega}{2k_B T}}$ in the integrand of the noise kernel. As shown in Figure \ref{fig:norm_bcf}(a), when $T$ increases, the noise kernel grows larger, leading to greater fluctuations. Consequently, the perturbation to the influence functional, which accounts for both fluctuation and dissipation, requires higher-order corrections due to the larger fluctuation terms. In the high temperature limit ($T \to \infty$), the time-convolutionless (TCL2n) approach becomes impractical, as it requires an infinite number of perturbation terms. It should also be noted that our coupling strength is relatively strong compared to the norm ratio between the fourth-order generator and the second-order generator. A lower coupling strength would result in a higher temperature range for the validity of TCL4.

\section{Summary}
Benchmarking of the fourth-order time-convolutionless (TCL4) master equation highlights its ability to improve accuracy over the second-order TCL (TCL2) master equation, particularly for population and coherence dynamics in the low-temperature regime. TCL4 effectively reduces positivity violations seen in TCL2, which arise from the negative noise kernel at early times. In this regime, TCL4 demonstrates strong agreement with exact numerical methods, offering a computationally efficient alternative for capturing non-Markovian dynamics in open quantum systems.
At higher temperatures and biases, the reliability of TCL4 decreases as fluctuations in the noise kernel grow and higher-order corrections become necessary. While TCL4 maintains better accuracy than TCL2 under intermediate conditions, performance degrades under high-temperature and high-bias conditions, often converging with or underperforming TCL2. These limitations align with the perturbative nature of TCL4, where the norm ratio of higher-order generators indicates the breakdown of the perturbative expansion in these parameter spaces.

Comparisons with exact numerical methods further emphasize the advantages and limitations of TCL4. TEMPO provides exact results but suffers from exponential scaling in computational complexity, making it impractical for long simulation times or large systems. HEOM offers robust performance at moderate to high temperatures but struggles at low temperatures due to the extensive basis required for representing the bath correlation function. In contrast, TCL4 delivers consistent computational efficiency across all regimes, offering a viable alternative in parameter spaces where exact methods are computationally prohibitive.

\begin{acknowledgments}
The authors express their gratitude to the School of Physics at the Georgia Institute of Technology for their invaluable support, which included access to computational resources and seed funding that significantly contributed to the success of this project.
\end{acknowledgments}

\section*{Data Availability Statement}
The data that support the findings of this study are available from the corresponding author upon reasonable request.

\appendix
\counterwithin{figure}{section}
\begin{widetext}
\section{The declining relexation rate with rising temperature}\label{appendix: relexation}
In the main text, we have seen an effect that the relaxation rate decreases with temperature, which is usually not the case in the weak coupling regime. A higher temperature means a higher spectral density and a stronger relaxation rate is expected, but here the parameters are near the edge of the weak coupling regime, which can change the temperature dependence of the relaxation rate.

The coupling operator $A$ in the eigenbasis is transformed by the eigenvector
\begin{equation}
    A'=\frac{1}{2} V'\sigma_z V=\frac{1}{2}(\sin{\theta}\sigma_z+\cos{\theta}\sigma_x)
\end{equation}

The Bloch-Redfield dissipator in the Hilbert-Schmidt space is 
\begin{equation}
    D_{BR}(t)=\Lambda(t)^\star\otimes A+A^\star \otimes \Lambda(t)-\mathbb{I} \otimes A\Lambda(t)-A^\star\Lambda^\star (t)\otimes \mathbb{I},
\end{equation}
where the operator is element-wise product of $A$ and $\Gamma$,
\begin{equation}
    \Lambda=A.*\Gamma^T=\Gamma_0 \sin{\theta}\sigma_z+\Gamma_{\Delta}\cos{\theta}\sigma_{+}+\Gamma_{-\Delta}\cos{\theta}\sigma_{-}.
\end{equation}
and then 
\begin{equation}
\begin{aligned}
    &D_{BR}= \frac{1}{2}\\&
    \begin{bmatrix}
        -J_{-\Delta}\cos{\theta}^2&J_0\sin{\theta}\cos{\theta}& J_0\sin{\theta}\cos{\theta}& J_{\Delta}\cos{\theta}^2\\
        \left(J_{-\Delta}+i(S_{-\Delta}-S_0)\right)\frac{\sin{\theta}\cos{\theta}}{2}& -2J_0\sin{\theta}^2-(\Gamma_{\Delta}+\Gamma_{-\Delta}^{\star})\frac{\cos{\theta}^2}{2}&(\Gamma_{\Delta}^{\star}+\Gamma_{-\Delta})\frac{\cos{\theta}^2}&-\frac{\sin{\theta}\cos{\theta}}{2}\left(J_{\Delta}+i(S_0-S_\Delta)\right)\\
        \left(J_{-\Delta}+i(S_{-\Delta}-S_0)\right)\frac{\sin{\theta}\cos{\theta}}{2}&
        (\Gamma_{\Delta}+\Gamma_{-\Delta}^\star)\cos{\theta}^2/2&-2J_0\sin{\theta}^2-(\Gamma_{\Delta}^{\star}+\Gamma_{-\Delta})\frac{\cos{\theta}^2}{2}&
        -\frac{\sin{\theta}\cos{\theta}}{2}\left(J_{-\Delta}+i(S_{-\Delta}-S_0)\right)\\
        J_{-\Delta}\cos{\theta}^2 &-J_0\sin{\theta}\cos{\theta}&-J_0\sin{\theta}\cos{\theta}& -J_\Delta\cos{\theta}^2   
    \end{bmatrix} 
\end{aligned}
\end{equation}
and the free evolution is 
\begin{equation}
    D_{f}=\begin{bmatrix}
0 & 0 & 0 & 0\\
0&-i\Delta&0&0\\
0&0&i\Delta&0\\
0&0&0&0
\end{bmatrix}.
\end{equation}
Combined with A(4) and A(5), the second order Redfield tensor is 
\begin{equation}
\mathcal{L}^2=D_f+D_{BR}
\end{equation}
We apply a unitary transformation to the Bloch-Redfield tensor
\begin{equation}
    U=\frac{1}{\sqrt{2}}\begin{bmatrix}
1 & 0 & 0 & 1\\
0&1&i&0\\
0&1&-i&0\\
1&0&0&-1
\end{bmatrix}
\end{equation}
which transform it into the Bloch basis, with real vector $|\rho>=\frac{1}{2}[1,n_x,n_y,n_z]^{\dagger}$, corresponding to the 
column vector of Pauli matrices $\mathbb{I}, \sigma_x, \sigma_y, \sigma_z$. Then,
\begin{equation}
D_{BR}'=  U^\dagger D_{BR} U  =\begin{bmatrix}
0 & 0 & 0 & 0\\
-(J_\Delta-J_{-\Delta})\frac{\sin\theta\cos\theta}{2}  & -J_0\sin^2\theta & 0 & (J_\Delta+J_{-\Delta})\frac{\sin\theta\cos\theta}{2}\\
[S_\Delta+S_{-\Delta}-2S_0]\frac{\sin\theta\cos\theta}{2} & 
[S_{-\Delta}-S_{\Delta}]\frac{\cos^2\theta}{2}
 
& -J_0\sin^2\theta- [J_\Delta+J_{-\Delta}]\frac{\cos^2\theta}{2}& [S_{-\Delta}-S_\Delta]\frac{\sin\theta\cos\theta}{2}\\
[J_\Delta-J_{-\Delta}]\frac{\cos^2\theta}{2} & J_0\sin\theta\cos\theta & 0 & -[J_\Delta+J_{-\Delta}]\frac{\cos^2\theta}{2}
\end{bmatrix}
\end{equation}
and the unitary evolution 
\begin{equation}
D_{f}'=U^\dagger D_{f} U =\begin{bmatrix}
0 & 0 & 0 & 0\\
0&0&\Delta&0\\
0&-\Delta&0&0\\
0&0&0&0
\end{bmatrix}
\end{equation}
The eigenvalues of $D_f+D_{BR}$ are composed of one relaxation rate and two dephasing rates, and the unitary transformation does not change the eigenvalue. Solving the eigenvalue problem of $D_{U}'+D_{BR}'$ gives us the relexation rate and dephasing rates but will be simpler since the first row zero. 

As the result of a random $\theta$ may be complicated, the solution in the case of zero bias $\theta=0$ is simple,
\begin{equation}
    \mathcal{L}^{2'}=  U^{\dagger} \mathcal{L}^{2} U  =\begin{bmatrix}
0 & 0 & 0 & 0\\
0  & 0 & \Delta & 0\\
0 & 
[S_{-\Delta}-S_{\Delta}]-\Delta
& - [J_\Delta+J_{-\Delta}]& 0\\
[J_\Delta-J_{-\Delta}] & 0 & 0 & -[J_\Delta+J_{-\Delta}]
\end{bmatrix}
\end{equation}
The cubic equation,
\begin{equation}
    \lambda^3-(J_{-\Delta}+J_{\Delta})\lambda^2+\left[J_{\Delta}+J_{-\Delta}+\Delta (S_{-\Delta}-S_{\Delta}-\Delta)\right]\lambda-(J_{\Delta}+J_{-\Delta})^2=0
\end{equation}
yields three eigenvalues $\lambda=-J_{-\Delta }-J_{\Delta }, \frac{1}{2} \left(-J_{-\Delta }-J_{\Delta }\pm \sqrt{\left(J_{-\Delta }+J_{\Delta }\right){}^2-4
   \left(\Delta ^2-\Delta  S_{-\Delta }+\Delta  S_{\Delta }\right)}\right)$. In the case $(J_{-\Delta }+J_{\Delta }){}^2<4
   \left(\Delta ^2-\Delta  S_{-\Delta }+\Delta  S_{\Delta }\right)$, the relaxation rate solely depends on the real part of spectral density $(J_{\Delta}+J_{-\Delta})$, which increases with rising temperature. However, if $(J_{-\Delta }+J_{\Delta }){}^2>4
   \left(\Delta ^2-\Delta  S_{-\Delta }+\Delta  S_{\Delta }\right)$, then the dynamics is overdamped\cite{breuer2002theory} and relexation rate can decrease with increasing temperature.
   
   In the case of \(\theta > 0\), the term \((S_{-\Delta} - S_{\Delta})\) is automatically incorporated into the real part of dephasing, which does not increase monotonically with temperature. Figure (\ref{fig:relaxation rate}) measures the relaxation rate for TEMPO and TCLs through curve fitting. We observe that the relaxation rate is at maximum at \(T = 2\Omega\) and then quickly drops across the higher temperatures hinders relexation.

\begin{figure}
    \centering
    \includegraphics[width=0.5\linewidth]{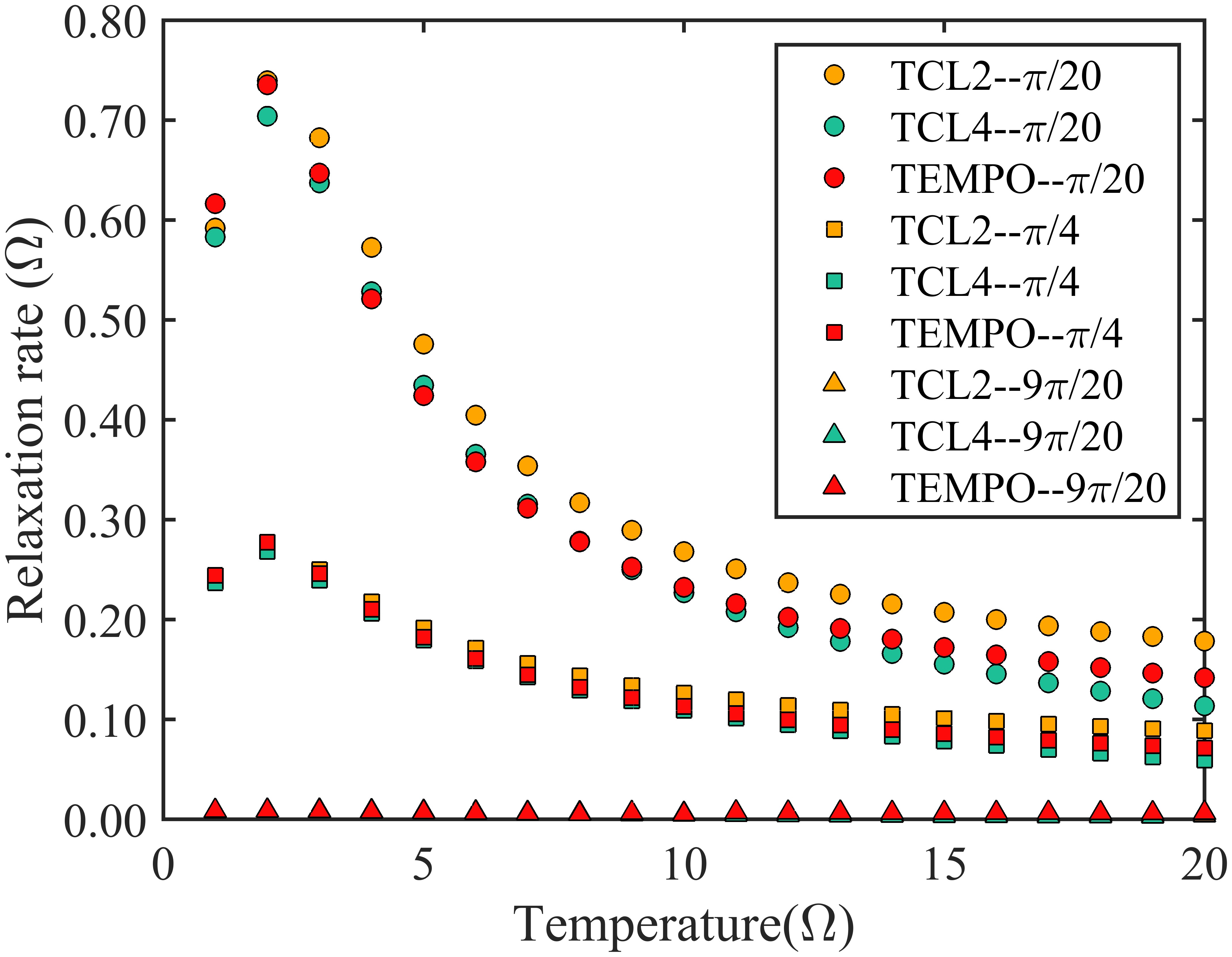}
    \caption{Relxation Rates at different bias and temperatures. Data measured by fitting the population using an exponential decay function $ae^{-\lambda t}+b$, where $\lambda$ is the relexation rate.  }
    \label{fig:relaxation rate}
\end{figure}
\end{widetext}

\section{Computation effort of TCL4}
The BCF is calculated from Eqn. \ref{eqn:bcf} at different temperatures 
\begin{equation}
    C(t)=\frac{1}{\pi}\int_{-\infty}^{\infty}d\omega'  e^{i\omega' t}\frac{J(\omega')}{e^{\beta\omega}-1}\\,
\end{equation}
with $J(\omega') = e^{-\beta \omega'} J(-\omega')$ for $\omega' < 0$, only the real part of the integrand has a temperature dependence. This integration is further discretized using a uniform time step $dt$, resulting in a time series $t_j$, where $j$ is the index for the time points. The first and last time points obtained after the fast Fourier transform (FFT) are $t_{-N}$ and $t_N$, respectively, with $t_N = Ndt$. The maximum frequency and the frequency difference for discretization are given by $\omega_M = \pi / dt$ and $d\omega = \frac{\pi}{Ndt}$, respectively.
 
\begin{equation}
    C(t_j)=\frac{1}{\pi}\sum_{k=-N,...N}e^{i \pi \frac{ k }{Ndt} t_j} \frac{J(\frac{k}{N dt}\pi)}{e^{\beta \frac{\pi k}{Ndt}}-1}=\frac{1}{\pi}\sum_{k=-N,...N}e^{i \pi \frac{ k }{N} j} \frac{J(\frac{k}{N dt}\pi)}{e^{\beta \frac{\pi k}{Ndt}}-1}.
\end{equation}

The convergence of the FFT depends on $\omega_M$ and $d\omega$. $\omega_M$ should be large enough to ensure that the gap between $J(\omega)$ at $\omega_M$ and $-\omega_M$ is closed. On the other hand, $d\omega$ needs to be sufficiently small to achieve the high resolution required for the spectral density. Both of these requirements necessitate a large number of $N$ points for the FFT. While the FFT is extremely fast for obtaining the BCF over long timescales, the bottleneck lies in its memory requirements compared to direct integration, particularly when the spectral density decays slowly over $\omega$ or exhibits rapid variations at small $\omega$ resolutions. Given the spectral density used in the main text at different temperatures, the convergence of the BCF at various $t_N$ values is shown in Table 1.

\begin{table}[htbp]
\centering
\begin{tabular}{llllll}
\toprule
\textbf{$t_N=Ndt$} &\makebox[3em]{5e0}&\makebox[3em]{1e1}&\makebox[3em]{5e1}
&\makebox[3em]{5e2}&\makebox[3em]{5e3}\\
\midrule
\textbf{T ($\Omega$)} &&&&&\\
0.1 &1e-2&1e-3&1e-5&1e-7&1e-9\\
1 &1e-2&1e-3&1e-5&1e-6&1e-8\\
5 &1e-1&1e-2&1e-3&1e-5&1e-6\\
\bottomrule
\end{tabular}
\caption{BCF convergence for ohmic spectral density at different temperatures (exponential cutoff). The numbers in the table are the order of magnitude difference between different $t_N$ and BCF calculated at $t_N=5\times 10^5$. $\omega_c=10$. } 
\end{table}
The time-dependent spectral density is calculated from the correlation function through a direct integration from $0$ to $t_{\text{end}}$, as shown in Eqn. \ref{eqn:time_sd}. This requires that $t_N$ be larger than $t_{\text{end}}$.

Unlike the second-order TCL generator, which only requires the spectral density at the local time, TCL4 involves a convolution of the spectral density over all past times. For the Markovian master equation, one only needs to calculate the generator at a large time where the generator saturates. The computation of a single generator requires a time complexity of $O(M^3n)$, where $M$ is the dimension of the Bohr frequency and $n$ is the number of timesteps up to the end time. If $dt$ is fixed, the computational time scales linearly with the simulation time. 

For the non-Markovian case, the TCL4 generators must be calculated at each timestep, resulting in a time complexity of $O(M^3n(n-1))$. In comparison, the TEMPO method requires a time complexity of $O(n \chi^3 M^6 K)$ \cite{fux2024oqupy}, where $\chi$ is the bond dimension.

\section{Comparison of an exponential cutoff}

\begin{widetext}

\begin{figure}
\captionsetup{justification=raggedright,singlelinecheck=false}
    \centering
\includegraphics[width=\linewidth]{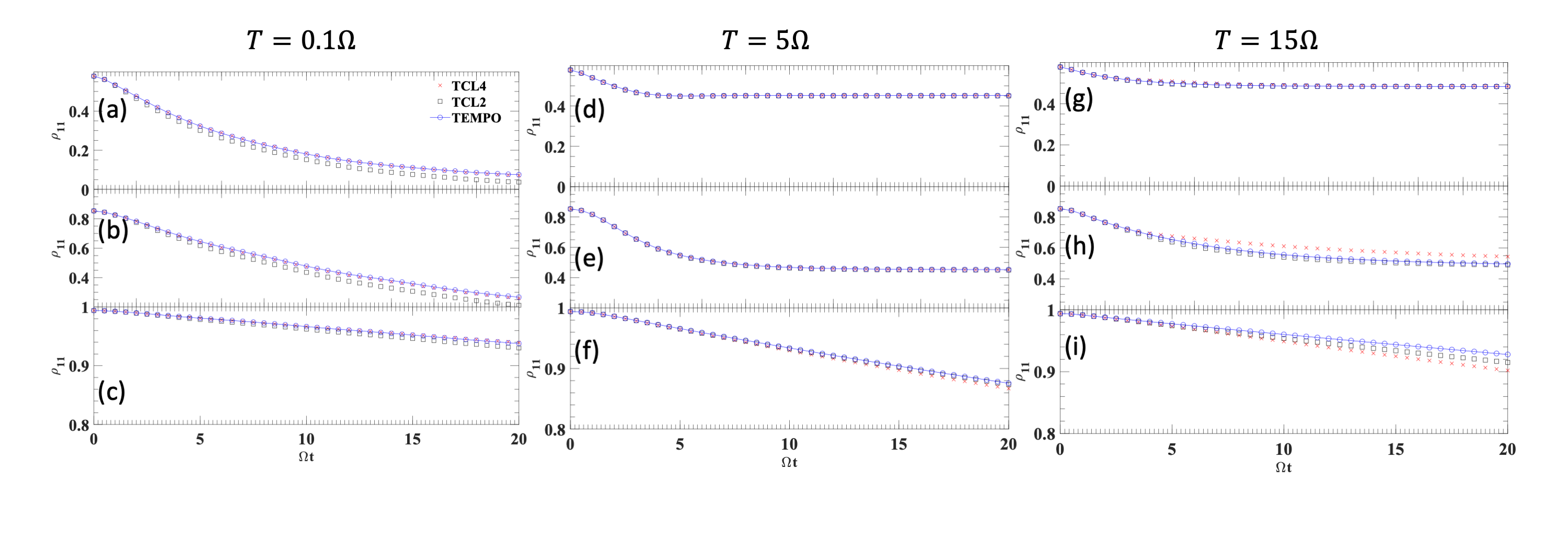}\caption{Population dynamics at different temperatures and bias of exponential cutoff $f_{exp}$. (a-c) $T=0.1\Omega$, $\theta=\pi/20$, $\pi/4$, $9\pi/20$. (d-e) $T=5\Omega$, $\theta=\pi/20$, $\pi/4$, $9\pi/20$.
    (g-i) $T=15\Omega$, $\theta=\pi/20$, $\pi/4$, $9\pi/20$. Model parameters: Simulation parameters: (TEMPO) $\Delta_t=0.01$, $\lambda_c=75$, $K=4000$. 
}    \label{fig:exp_pop}
\end{figure}

\begin{figure}
\captionsetup{justification=raggedright,singlelinecheck=false}
    \centering
\includegraphics[width=\linewidth]{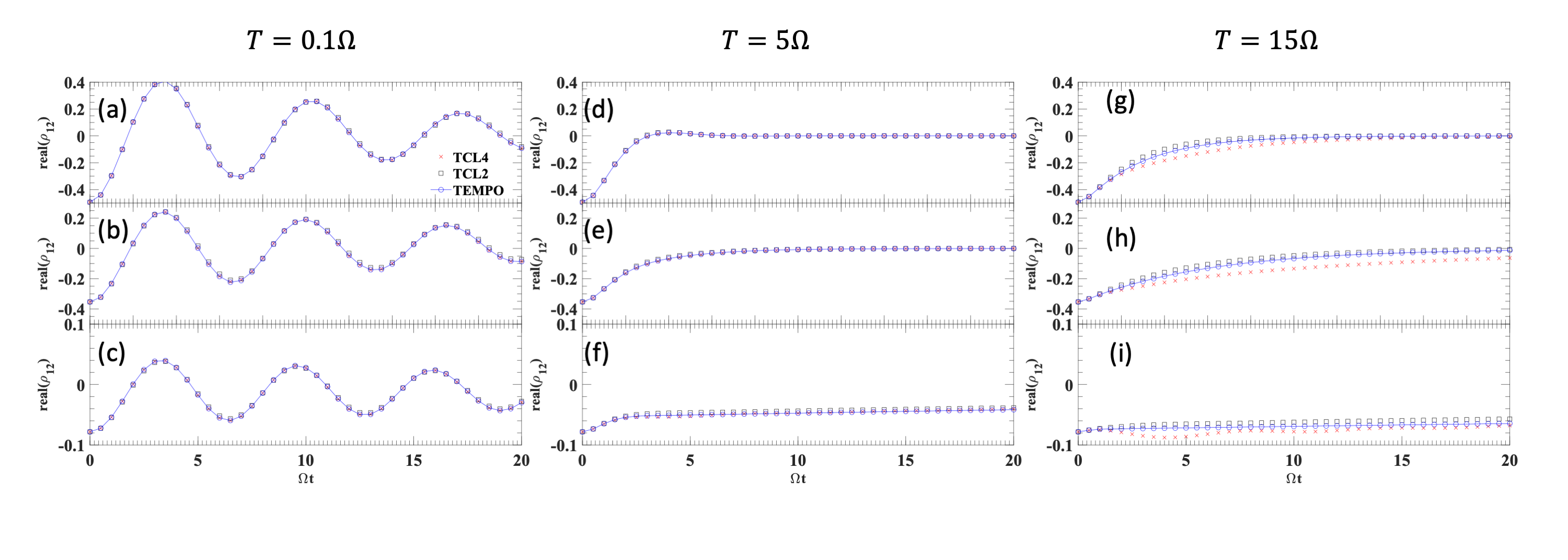}
    \caption{Coherence dynamics at different temperatures and bias of exponential cutoff $f_{exp}$. (a-c) $T=0.1\Omega$, $\theta=\pi/20$, $\pi/4$, $9\pi/20$. (d-e) $T=5\Omega$, $\theta=\pi/20$, $\pi/4$, $9\pi/20$.
    (g-i) $T=15\Omega$, $\theta=\pi/20$, $\pi/4$, $9\pi/20$. Model parameters: Simulation parameters: (TEMPO) $\Delta_t=0.01$, $\lambda_c=75$, $K=4000$.} \label{fig:exp_coh}
\end{figure}
\end{widetext}
\FloatBarrier
In complement of the Drude cutoff, here we present the comparison of the TCL2, TCL4 and TEMPO methods with exponential cutoff bath in Figures \ref{fig:exp_pop} and \ref{fig:exp_coh}, where HEOM cannot be used. $T=0.1 \Omega$, $5\Omega$, and $15 \Omega$ are used. A similar pattern of error is found in most regions. TCL4 manifests its advantage at low temperature and its inability to operate at high temperatures of both coherence and population. However, the difference is at high temperature and high dephasing area, where $T=15\Omega$ and $\theta=9\pi/20$, where TCL4 shows a signature of bouncing coherence.

\begin{figure}
    \centering
\captionsetup{justification=raggedright,singlelinecheck=false}
\includegraphics[width=0.7\linewidth]{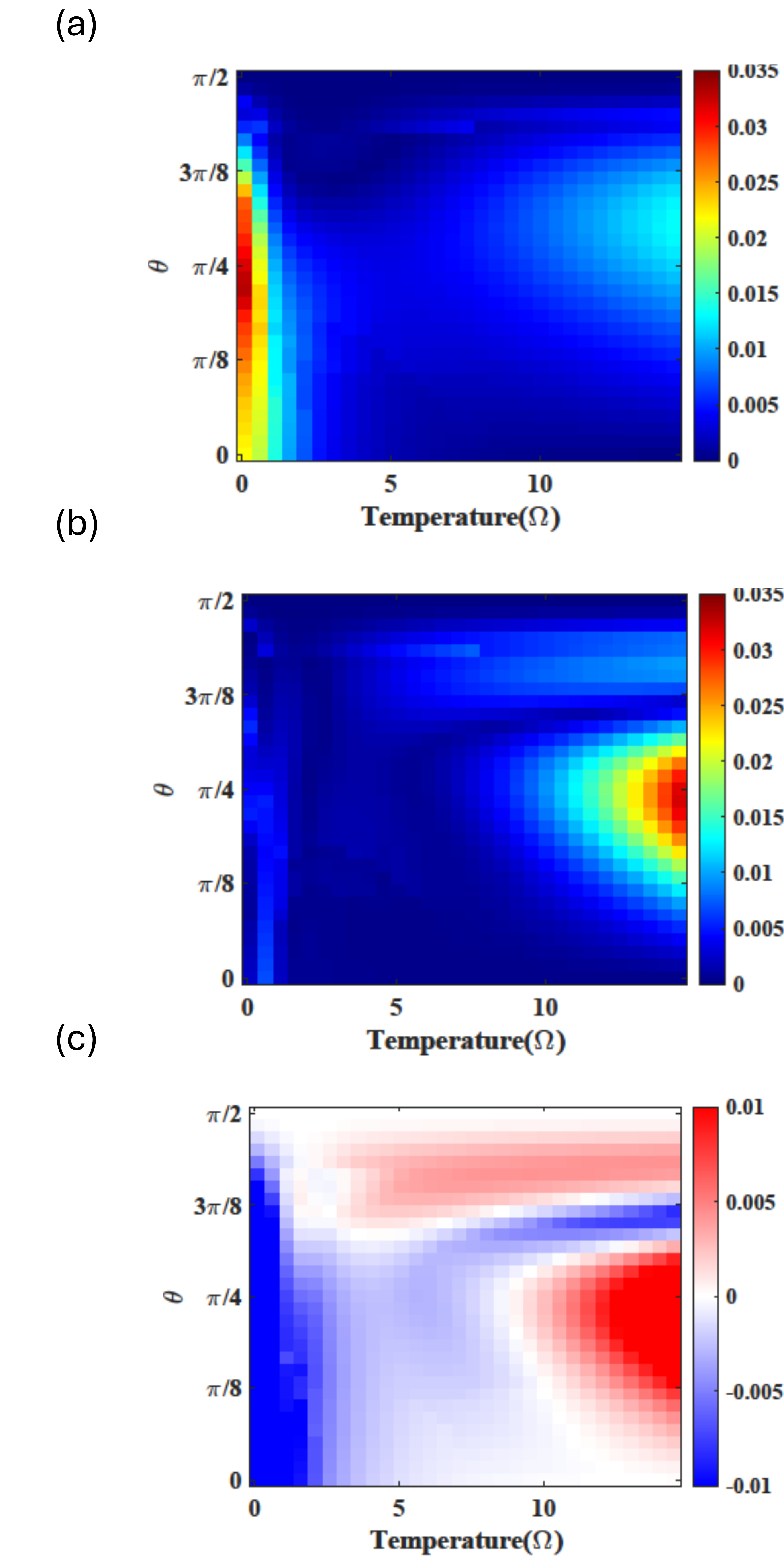}
    \caption{The time-averaged trace distance between TCLs and TEMPO of spectral density with exponential cutoff. (a) $\Delta_{TCL2, TEMPO}$. (b)$\Delta_{TCL4,TEMPO}$. (c)$\Delta_{TCL4,TEMPO}-\Delta_{TCL2,TEMPO}$. Simulation parameters, $\lambda^2=0.2$. $\omega_c=10$. Convergence parameters: (TEMPO) $T=1$, $\Delta_t=0.01$, $\lambda_c=80$, $K=4000$. For other temperatures, $\Delta_t=0.01$, $\lambda_c=75$, $K=4000$. $t_{end}=15$  }
    \label{fig:exp_trace}
\end{figure}

Figs. \ref{fig:exp_trace} (a) and (b) show $\Delta$ between TCL2 and TEMPO, and between TCL4 and TEMPO. The diagram is more complicated than the one with Drude spectral density as one more region shows that the TCL4 is worse than TCL2. The low-temperature dominance of TCL4 and the high-temperature, low-bias region are trending similarly to a bath with a Drude cutoff.
In the case of pure dephasing ($\theta=\pi/2$), both TCL4 and TCL2 show no deviation from TEMPO, confirming the exactness of the second-order TCL equation for pure dephasing. However, as soon as the bias shifts downward from $\theta=\pi/2$, the equations of TCL2 and TCL4 cease to be exact and enter an error regime where TCL4 gains more deviation from the result of TEMPO. Fig. \ref{fig:exp_trace} (c) clearly manifests this area at temperatures above $4\Omega$, where TCL4 introduces more error in addition to the second-order generator due to coherence reoccurrence, meaning the coherence becomes bouncing along the exact result. 

\FloatBarrier

\bibliography{aipsamp}% Produces the bibliography via BibTeX.

\end{document}